\documentclass[preprint,showpacs,preprintnumbers,amsmath,amssymb,nofootinbib]{revtex4}

\usepackage{epsf}
\usepackage{graphicx}
\usepackage{dcolumn}
\usepackage{bm}

\def\be{\begin{equation}}
\def\ee{\end{equation}}
\def\bea{\begin{eqnarray}}
\def\eea{\end{eqnarray}}

\newlength{\dinwidth}
\newlength{\dinmargin}
\newcommand{\ba}{\begin{eqnarray}}
\newcommand{\ea}{\end{eqnarray}}

\def\ga{\mathrel{\mathpalette\fun >}}
\def\fun#1#2{\lower3.6pt\vbox{\baselineskip0pt\lineskip.9pt
        \ialign{$\mathsurround=0pt#1\hfill##\hfil$\crcr#2\crcr\sim\crcr}}}

\def\figsize3{10cm}
\def\figsize{8.5cm}

\begin{document}

\preprint{astro-ph/0409264}
\preprint{ SLAC-PUB-10697 }

\title{Current Observational Constraints on Cosmic Doomsday}
\author{Yun Wang$^1$, Jan Michael Kratochvil$^2$, Andrei Linde$^2$, and
Marina Shmakova$^3$}
\affiliation{$^1$Department of Physics \& Astronomy,
 Univ.\ of Oklahoma\\
 440 W Brooks St., Norman, OK 73019\\
  email: wang@nhn.ou.edu}
\affiliation{ {$^2$Department
  of Physics, Stanford University, Stanford, CA 94305-4060,
USA}    }
\affiliation{$^3$  SLAC,
 Stanford University, Stanford, CA 94309  }

\date{\today}
{\begin{abstract} In a broad class of dark energy models, the
universe may collapse within a finite time $t_c$. Here we study a
representative model of dark energy with a linear potential,
$V(\phi)=V_0 (1+ \alpha \phi)$. This model is the simplest
doomsday model, in which the universe collapses rather quickly
after it stops expanding. Observational data from type Ia
supernovae (SNe Ia), cosmic microwave background anisotropy (CMB),
and large scale structure (LSS) are complementary in constraining
dark energy models. Using the new SN Ia data (Riess sample), the
CMB data from WMAP, and the LSS data from 2dF, we find that the
collapse time of the universe is $t_c \ga 42$ (24) gigayears from
today at 68\% (95\%) confidence.
\end{abstract}}


 \maketitle

\section{Introduction}

The existence of dark energy \cite{Riess98,Perl99} is one of the most
significant cosmological discoveries ever made. The nature of this dark energy
has remained a mystery. Viable models of dark energy include a cosmological
constant, quintessence \cite{Banks,Linde,quintessence}, or modified Friedmann
equations \cite{modifiedgravity}. The only available model of dark energy based
on string theory describes a metastable cosmological constant with an
exponentially large decay time \cite{KKLT}. See \cite{reviews} for reviews with
more complete lists of references.

Although current observational data seem to favor a cosmological constant
\cite{Riess04,WangTegmark,Jassal,Huterer04}, there are two good reasons to continue dark
energy studies. First, the measured dark energy density\footnote{The dark energy
density is better constrained by data than the dark energy equation of state,
since it is more closely related to data.\cite{WangFreese04}}
as an arbitrary
function of time, $X(z) \equiv \rho_X(z)/\rho_X(0)$, has large uncertainties
\cite{WangMukherjee,WangTegmark,Daly04}, hence it does not rule out dark energy
models with $\rho_X(z)$ that do not vary greatly with time \cite{variousX}.
Second, even if $\rho_X(z)$ were measured to be a constant at high accuracy in
future observations, we still need to  find physically well motivated models
that give an effective cosmological constant with an incredibly small value
$\Lambda \sim 10^{-29}$ g/cm$^3$.

One of the first (and by far the simplest) models of dark energy, which we are
going to discuss in our paper, was proposed in \cite{Linde} in order to address
the cosmological constant problem. In this model the cosmological constant was
replaced by the energy density of a slowly varying scalar field $\phi$ with the
linear effective potential
\be V(\phi)=V_0 (1+ \alpha \phi). \label{eq:V(phi)} \ee
Here we use the units $M_p=(8\pi G)^{-1/2}=1$.  If the slope of the potential
is sufficiently small, $\alpha V_0\lesssim 10^{-120}$ in Planck units,  the
field $\phi$ practically does not change during the last $10^{10}$ years, its
kinetic energy is very small, so its total potential energy $V(\phi)$ acts
nearly like a cosmological constant. The anomalous flatness of the effective
potential in this scenario is a standard feature of most of the models of dark
energy \cite{Banks,Linde,quintessence}.

Even though the energy density of the field $\phi$ in this model practically
does not change at the present time, it changed substantially during inflation.
Since $\phi$ is a massless field, it experienced quantum jumps with the
amplitude $H/2\pi$ during each time interval $H^{-1}$. These jumps moved the field
$\phi$ in all possible directions. As a result,  the universe became divided
into an infinitely large number of exponentially large `universes' containing
all possible values of the field $\phi$, and, consequently, all possible values
of the effective cosmological constant $\Lambda(\phi) =V(\phi)$. This quantity
may range from $-M_p^4$ to $+M_p^4$ in different parts of the universe, but we
can live only in the `universes' with $|\Lambda| \lesssim 10^{-28}$ g/cm$^3$.

Indeed, if $\Lambda \lesssim -10^{-28}$ g/cm$^3$, the universe collapses within
a time much smaller than the present age of the universe $\sim 10^{10}$ years
\cite{Linde:ir,Linde, Barrow}. On the other hand, if $\Lambda \gg 10^{-28}$
g/cm$^3$, the universe at present would expand exponentially fast, energy
density of matter would be exponentially small, and life as we know it would be
impossible \cite{Linde:ir,Linde}. This means that we can live only in those
parts of the universe where the cosmological constant does not differ too much
from its presently observed value $|\Lambda| \sim 10^{-29}$ g/cm$^3$. This
approach constituted the basis for many subsequent attempts to solve the
cosmological constant problem using the anthropic principle in inflationary
cosmology \cite{Barrow,Weinberg87,GV,Don,Kallosh}.

There are several other models of a slowly rolling scalar field which may allow
one to solve the cosmological constant problem. For example, one may consider
models with quadratic potential $m^2\phi^2/2 -\Lambda$ \cite{Banks,GV}, with
potentials $V_0 e^{\alpha\phi}-\Lambda$ \cite{exp}, potentials unbounded from
below in $N = 8$ supergravity \cite{Kallosh}, or potentials depending on two
different fields $\alpha V_0 \phi +\beta\xi^n$ \cite{Garriga:2003hj}. In all of
these models the anthropic solution of the cosmological constant problem is
based on the assumption that the effective potential continuously changes in a
sufficiently broad interval near $V =0$. In all of these models the scalar
fields eventually roll down to negative values of $V$,  and the universe
experiences global collapse, just as the universe with a negative cosmological
constant. A detailed description of this process can be found in
\cite{negative}.

 Thus, in all of the models
mentioned above, {\it the  anthropic solution of the cosmological constant
problem goes hand in hand with the  prediction of the global collapse of the
universe}.\footnote{One can avoid this conclusion in the models where the
cosmological constant can take an exponentially large number of {\it discrete}
values near $V=0$ \cite{BP,Susskind,Douglas}. In the string theory based models
of this type the universe may either collapse \cite{Kallosh:2004yh} or 
spontaneously decompactify and become 10-dimensional \cite{KKLT}.} This
 makes it especially interesting whether there could be
any observational evidence in favor of these models, and what kind of bounds on
the lifetime of the universe one could obtain by cosmological observations.

Talking about the future observations,  even if the results to be obtained by
JDEM and Planck satellites will favor the simplest cosmological constant model,
they will not necessarily imply that the universe is going to expand forever.
They will only place some model-dependent limits on the possible time of the
doomsday. For example, in the context of the simplest linear model
(\ref{eq:V(phi)}), such data would only imply that the collapse of the universe
cannot happen earlier than $t_c \sim 40$ gigayears from now, at the 95\%
confidence level \cite{Kalloshetal}.

The actual data to be obtained many years from now may alter these estimates.
It is important to understand what we can say about the possible doomsday time
on the basis of the presently available data. There were several attempts to
study this question \cite{exp,Kalloshetal,Alam:2003rw,Riess04,WangTegmark}, but
the results of these investigations appear to be strongly model-dependent. Our
approach will be similar to that of Ref. \cite{Riess04}, where the constraint
on the doomsday time $t_c \gtrsim 30$ Gyrs was obtained for  models with the
potential $V_0 e^{\phi/2}-\Lambda$. However, this constraint would change
significantly if one considered a more general function $V_0
e^{\alpha\phi}-\Lambda$ depending on three parameters, $\alpha$, $V_0$, and
$\Lambda$, in addition to the initial value of the field $\phi$.

In this paper we are going to study the observational constraints on the
lifetime of the universe in the simplest but rather general class of models
where the effective potential with respect to the slowly rolling field $\phi$
can be represented by the linear function (\ref{eq:V(phi)}) in the
cosmologically interesting interval $|V(\phi)| \lesssim 10^{-28}$ g/cm$^3$
\cite{Linde}. An advantage of this model (see also \cite{DimThom}) is that it
depends only on one parameter $\alpha V_0$. This can be made manifest by a
shift $\phi \to \phi - V_0/\alpha$ which reduces the potential (1) to an even
simpler expression $\alpha V_0 \phi$. This will allow us to obtain more
definite constraints on the doomsday $t_c$.

We will use the Markov Chain Monte Carlo (MCMC) technique, which will allow us
to compare observational data with the model (\ref{eq:V(phi)}) with different
values of $\alpha$. Even for this simple model,  we still need an additional
input from the theory. The simplest and perhaps the most reasonable approach is
to say that since we do not know anything about the origin of the parameter
$\alpha$, we will consider all values of $\alpha$ being equally probable, and
use flat prior probability distribution $p(\alpha) = const$ as an input in the
MCMC calculations. But various theoretical arguments may give different prior
probability distributions $p(\alpha)$, which may affect the results. For
example, one may argue that if there is some reason why the parameter $\alpha
V_0$ in (\ref{eq:V(phi)}) is smaller than $10^{-120}$, in Planck units, then
perhaps it is many orders of magnitude smaller than $10^{-120}$ \cite{GVdark}.
This is a very strong argument which suggests that $p(\alpha)$ may be strongly
peaked at small $\alpha$, in which case the theory will have almost exactly the
same observational consequences as the simple cosmological constant model. On
the other hand, instead of the model (\ref{eq:V(phi)}) one may consider models
with potentials $\alpha V_0 \phi +\beta\xi^n$. During inflation, each of these
two fields will fluctuate, so that in different parts of the universe with $|V|
\lesssim 10^{-28}$ g/cm$^3$ one will have different combinations of the fields
$\phi$ and $\xi$, and the slope of the effective potential in the steepest
descent direction will also take different values. Such models in the
observationally important region $|V| \lesssim 10^{-28}$ g/cm$^3$ behave just
as the model with a constant slope of the effective potential (\ref{eq:V(phi)}) with respect to a combination of the fields $\phi$ and $\chi$, but the prior probability distribution
to live in a part of the universe with a given effective slope $\alpha V_0$
will depend on the parameter $n$ \cite{Garriga:2003hj}.

That is why in our calculations we will examine the linear model
(\ref{eq:V(phi)}) using the simplest prior probability $p(\alpha) = const$, as
well as two other possibilities, $p(\alpha) \propto \alpha$ and $p(\alpha)
\propto \alpha^{-1/2}$. Whereas the resulting constraints on the lower bound on
the doomsday time $t_c$ appears not to be very sensitive to the choice of
$p(\alpha)$, one should keep in mind that, in general, predictions of the fate
of the universe depend not only on the choice of the potential, but also on the
prior probability for the parameters of the model. This does not make the results of the MCMC calculations ill defined, it just means that the definition of a physical model of dark energy should include some idea about the probability of the parameters of the model, as well as the probability of initial conditions for the scalar field. In the absence of any {\it a priori} idea, it is reasonable to use a flat prior for these parameters.

In Sec. II of our paper we will describe our analysis technique and the data
used. We present our results in Sec.\ III. In Sec.\ IV we will make some
methodological comments, comparing the results of the MCMC calculations with
the Fisher matrix approach. Sec.\ V contains a critical discussion of the
possibility to obtain model-independent constraints on the doomsday time. We
summarize our results and discuss their implications in Sec.\ VI.

\section{Method}

\subsection{Evolving the scalar field}

Following Ref.\cite{Kalloshetal}, we assume a flat universe,
$\Omega_{tot}=\Omega_m + \Omega_{\phi}=1$, which is consistent with current
observational data \cite{WMAP,Tegmark03}. Note that since the expansion rate of
the universe today, $H_0$, depends on $\Omega_m$, $V_0$ and $\alpha$, the only
independent parameters in this model are ($\Omega_m$, $\alpha$, $H_0$).

The usual equations of motion are
\ba
&& \ddot{\phi} + 3 H(a) \dot{\phi} + \frac{\partial V}{\partial\phi}=0, \\
& & \frac{\ddot{a}}{a} = \frac{8\pi G}{3} \left[ V(\phi) - \dot{\phi}^2 -
\frac{1}{2} \frac{\rho_m^0}{a^3} \right],
\ea
where $a$ is the cosmic scale factor, $\rho_m^0$ is the matter density
today, and the dots denote derivatives with respect to the cosmic time $t$.
The Hubble parameter is
\be
H^2(a) \equiv \left( \frac{\dot{a}}{a} \right)^2= \frac{8\pi G}{3}
\left[ \frac{\rho_m^0}{a^3} +
\frac{\dot{\phi}^2}{2} + V_0 (1+\alpha \phi) \right],
\label{eq:H^2}
\ee

It is not straightforward to solve the usual equations of motion for
a given set of ($\Omega_m$, $\alpha$, $H_0$), since $\Omega_m$
depends on $H_0$, which in turn is only known once we have
solved the equations for $a(t)$ and found its derivative at $a=1$.

We rewrite the equations of motion as \ba \label{eq:efm}
&&\frac{d^2\overline{\phi}}{d\tau^2} + 3 E_*(z)\,
\frac{d\overline{\phi}}{d\tau}
+ \overline{\alpha} =0, \\
&& \frac{1}{a} \frac{d^2a}{d\tau^2}=1+ \overline{\alpha}\overline{\phi}
- \left( \frac{d\overline{\phi}}{d\tau} \right)^2 -
\frac{1}{2} \frac{\Omega_m^*}{a^3},\\
&& \left( \frac{1}{a} \frac{da}{a\tau} \right)^2 =E_*^2(z)
\equiv 
\frac{\Omega_m^*}{a^3}+ \frac{1}{2} \left(\frac{d\overline{\phi}}{d\tau}
\right)^2 +1 + \overline{\alpha} \overline{\phi},
\label{eq:E_*}
\ea
where
\be
\tau \equiv \sqrt{ \frac{8\pi G V_0}{3}} \, t,
\hskip 1cm \overline{\phi} \equiv\sqrt{ \frac{8\pi G }{3}} \, \phi,
\hskip 1cm \overline{\alpha} \equiv\sqrt{ \frac{3}{8\pi G }}\, \alpha,\hskip 1cm
\Omega_m^* \equiv \frac{\rho_m^0}{V_0}.
\ee
Note that the equations of motion (Eqs.[\ref{eq:efm}]-[\ref{eq:E_*}])
only depend on ($\Omega_m^*$, $\overline{\alpha}$).

At $\tau \rightarrow 0$, $a  \rightarrow 0$,
we take $\phi=d\phi/dt=0$ \cite{Kalloshetal}. Evolving the
equations of motion to $a=1$ ($z=0$) gives us
\be
H_0 = \sqrt{ \frac{8\pi G V_0}{3}} \, E_*(z=0),
\hskip 1cm \Omega_m = \frac{\Omega_m^*}{E_*^2(z=0)}.
\ee

\subsection{Prior distributions of parameters}

In the linear model of Eq.(\ref{eq:V(phi)}), the expansion rate of the universe
depends on the parameters ($\rho_m^0$, $V_0$, $\alpha$) (see
Eq.(\ref{eq:H^2})). We assume uniform priors for $\rho_m^0$ and $V_0$. It
can be shown that choosing uniform prior for $V_0$ is equivalent to
choosing uniform prior for the initial value of the scalar field $\phi$ after
inflation, which is the standard assumption made in
\cite{Linde,Garriga:2003hj}. For $\alpha$, we consider three different priors
(see Ref.\cite{Garriga:2003hj} for justifications): \be P(\alpha) \propto 1;
\hskip 1cm P(\alpha) \propto \alpha^{-0.5}; \hskip 1cm P(\alpha) \propto \alpha
\ee which correspond to $n=2,3,3/2$ in Eq.(29) of Ref.\cite{Garriga:2003hj}.

Using the Bayes' Theorem, we can take the prior in $\alpha$
into consideration by replacing the usual $\chi^2$ with
\be
\tilde{\chi}^2= \chi^2 - 2\,\ln P(\alpha).
\ee

\subsection{Data used}

The measured distance-redshift relations of type Ia supernovae (SNe Ia)
provide the foundation for testing dark energy models.
In a flat Universe, the dimensionless luminosity distance
\be
d_L(z) = c H_0^{-1}\, (1+z) \Gamma(z), \hskip 1cm \Gamma(z)=\int_0^z dz'/E(z'),
\ee
where
\be
E(z)= E_*(z)/E_*(z=0)
\label{eq:E(z)}
\ee
with $E_*(z)$ given by Eq.(\ref{eq:E_*}).

The observational data we use in this paper are the same as
in \cite{WangTegmark}.
We use the ``gold'' set of 157 SNe Ia (the Riess sample) published in \cite{Riess04}
and analyze it using flux-averaging statistics \cite{Wang00b,WangMukherjee}
to reduce the bias due to weak gravitational lensing by intervening matter
\cite{lensing}.
A Fortran code that uses flux-averaging statistics
to compute the likelihood of an arbitrary
dark energy model (given the SN Ia data from \cite{Riess04})
can be found at \textit{http://www.nhn.ou.edu/$\sim$wang/SNcode/}
\cite{Wang00b,WangMukherjee,WangTegmark}.

We only use CMB and LSS data that are not sensitive to the
assumptions made about dark energy \cite{Knop,WangTegmark}.

The only CMB data we use is the measurement of the CMB shift
parameter \cite{Bond97},
\be
R\equiv \Omega_m^{1/2}\Gamma(z_{CMB})=1.716\pm 0.062
\ee
from CMB (WMAP, CBI, ACBAR) \cite{WMAP,CMB},
where $z_{CMB}=1089$.

The only large-scale structure information we use is
the linear growth rate $f(z_{2dF})=0.51\pm 0.11$
measured by the 2dF galaxy redshift survey (2dFGRS) \cite{Hawkins03,Verde02},
where $z_{2dF}\simeq 0.15$ is the effective redshift of this survey.
Since $f = \beta * b_1$, where
$\beta$ is the redshift distortion parameter measured from the
ratio of the redshift-space correlation function to the
real-space correlation function [see Eq.(17) in \cite{Hawkins03}],
and $b_1$ is the bias factor [square root of the ratio of galaxy power spectrum
and mass power spectrum]. Since both correlation function and power
spectrum are statistical descriptions of galaxy survey data, they can be
extracted from data without making specific assumptions about 
dark energy.
Note that on the other hand, the theoretical prediction
of the linear growth rate {\it does} depend on assumption
about dark energy, as well as cosmological parameters;
this is why we can use galaxy survey data to probe dark energy 
and constrain cosmological parameters.
For a given set of cosmological parameters and an assumed
dark energy density $\rho_X(z)$,
the linear growth rate $f\equiv (d\ln D/d\ln a)$ is determined 
by solving the equation for the linear growth rate $D$,
\be
D''(\tau) + 2E(z)D'(\tau) - {3\over 2}\Omega_m (1+z)^3D = 0,
\ee
where primes denote $d/d(H_0 t)$, and
$E(z)= \left[\Omega_m (1+z)^3 + (1-\Omega_m) \rho_X(z)/\rho_X(0)\right]^{1/2}$
for a flat universe. For the dark energy model studied in
this paper, $E(z)$ is given by Eqs.(\ref{eq:E(z)}) and (\ref{eq:E_*}).

We use the Markov Chain Monte Carlo (MCMC) technique in the likelihood analysis
(based on the MCMC engine of \cite{Lewis02}), and
obtain a few million samples of ($\Omega_m^*$, $\overline{\alpha}$, $H_0$).
This method samples from the full posterior distribution of the
parameters, and from these samples the marginalized posterior distributions
of the parameters (i.e., their probability distribution functions [pdf])
can be estimated. 
We use the method proposed by Gelman and Rubin 
to test for convergence \cite{Gelman92,Verde03}.
This method uses a convergence indicator 
\be
\hat{R}=\frac{ [(N-1)/N]W + B_n (1+1/M)}{W},
\ee
where $M$ is the number of chains (each with 2$N$ elements) starting at well-separated points 
in parameter space, W is the mean variance of the chains, and $B_n$ is
the variance between the chains. Convergence is achieved for 
$\hat{R} < 1.2$. We find that $\hat{R} < 1.01$ for our MCMC chains,
which assures us that convergence has been achieved.

\section{Results}
\label{results}

In models that lead to a Big Crunch, the Hubble parameter $H(z) \propto E(z)$
will decrease with time until $E(z)=0$ at $t=t_{turn}$, when the universe stops
expanding and starts to collapse. Fig.1 of Ref.\cite{Kalloshetal}  shows the
cosmic scale factor in five models (the linear potential model with five
different parameter choices). Clearly, the universe collapses rather quickly
after it stops expanding.

\begin{figure}[h]
\vskip-0.3cm \centerline{\epsffile{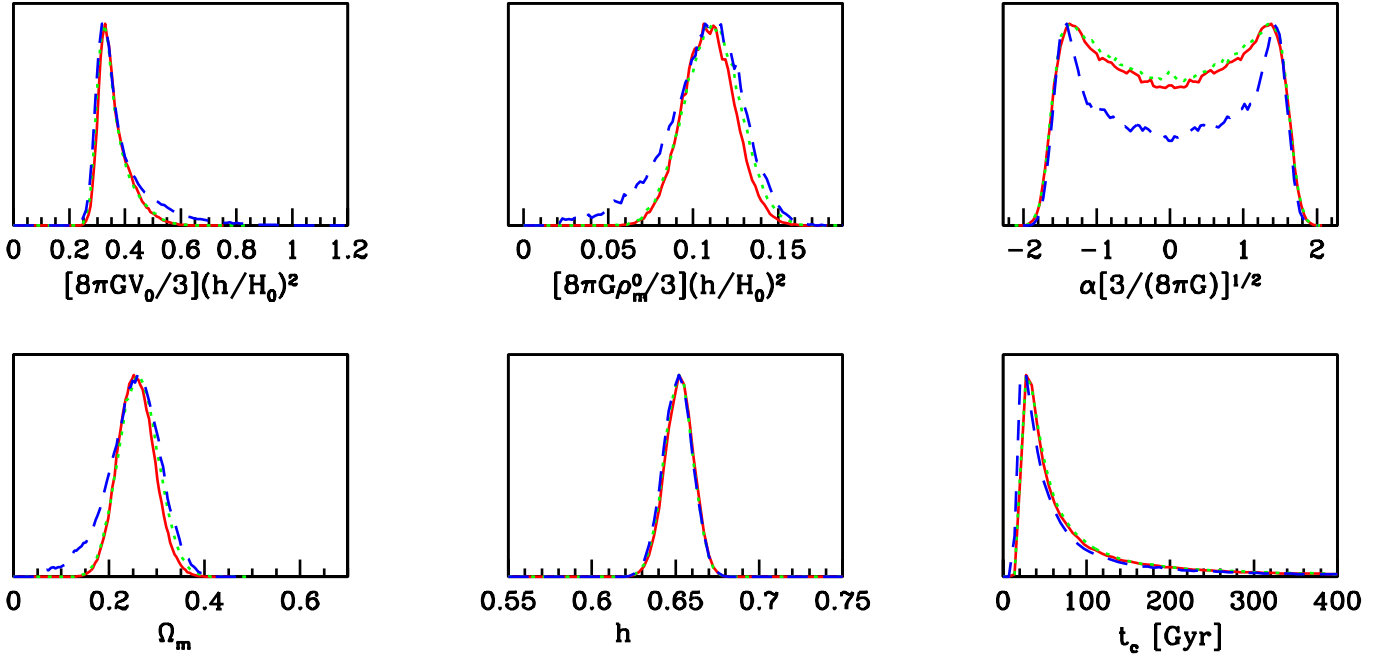}}
\vskip -10cm
\caption[1]{\label{pdf3set}\footnotesize%
The constraints on the linear model parameters from current SN data only (Riess
sample gold set, flux-averaged with $\Delta z=0.05$) (dashed lines), SN
together with CMB data (dotted lines), SN together with CMB and LSS data (solid
lines). Uniform priors on ($\rho_m^0$, $V_0$, $\alpha$) are assumed. The first
row shows the probability distribution functions of the set of independent
parameters $V_0$, $\rho_m^0$, and $\alpha$ The second row shows derived
parameters $\Omega_m$, $h$, and the time to collapse from today $t_c$. }
\end{figure}

Fig.\ref{pdf3set} shows the constraints on the linear model
from current observational data,
assuming uniform priors on ($\rho_m^0$, $V_0$, $\alpha$).
The first row of figures in
Fig.\ref{pdf3set} shows the probability distribution functions
of the set of independent parameters $V_0$, $\rho_m^0$, and $\alpha$.
The second row of figures in Fig.\ref{pdf3set} shows
derived parameters $\Omega_m$, $h$, and the time to collapse
from today $t_c$.
The dashed lines denote results using SN data only, the dotted
lines denote results using SN together with CMB data,
the solid lines denote results using SN together with CMB and LSS data.

Table 1 lists the mean and the 68\% and 95\% confidence ranges of
$\Omega_m$, $h$, $\alpha$, and $t_c$ from Fig.\ref{pdf3set}.

\newpage

\begin{center}
{\footnotesize
{Table 1\\
The constraints on the linear potential dark energy model
from current data}\\
\begin{tabular}{|l|l|l|l|}
\hline
\hline
& $\,$ SN only$^a$ & $\,$ SN plus CMB$^b$ & $\,$ SN plus CMB and LSS$^c$ \\
\hline
$\Omega_m$ & $.25 \,[.19,.30] [.12,.35]\,$  $\,$
& $.26 \,[.22,.30] [.18,.35]\,$  $\,$
& $.26 \,[.22,.30] [.19,.33]\,$  $\,$ \\\hline
$h^d$ & $.651 \,[.642,.660] [.634,.668]\,$  $\,$
& $.651 \,[.643,.660] [.635,.668]\,$
& $.652 \,[.644,.661] [.635,.668]\,$ \\\hline
$\alpha \left( \frac{3}{8\pi G} \right)^{1/2}$
& $0.00 \,[-1.30,1.30] [-1.60,1.60]\,$  $\,$
& $0.00\, [-1.20,1.20] [-1.60,1.60] \,$
& $0.00 \,[-1.20,1.20] [-1.60,1.60] \,$  \\\hline
$t_c$/[Gyr]$^e$ & $\,$ 56.01 [35.75, ---] [19.18, ---] $\,$
& $\,$ 67.48 [42.95, ---] [23.61, ---]
& $\,$ 65.66 [41.92, ---] [23.74, ---]\\
  & \hskip 1cm [---, 108.18] [---, 2704.01]
  & \hskip 1cm [---, 134.34] [---, 4095.18]
  & \hskip 1cm [---, 130.52] [---, 3652.01]
\\\hline
$\chi^2_{min}$/$N_{dof}^f$
& $\,$ 25.98/23 $\,$
& $\,$ 26.04 /24 & $\,$ 26.16/25 \\
\hline
\hline
\end{tabular}
}
\end{center}
{\footnotesize{$^a$ Riess sample gold set (157 SNe Ia), flux-averaged with ${\Delta}z=.05$.\\
$^b$ CMB shift parameter $R\equiv \Omega_m^{1/2}\Gamma(z_{CMB})=1.716\pm 0.062$.\\
$^c$ The linear growth rate $f(z_{2dF})=0.51\pm 0.11$.\\
$^d$ Statistical error only, not including the contribution
from the much larger SN Ia absolute magnitude error of $\sigma_h^{int}\simeq 0.05$
\cite{WangMukherjee}.\\
$^e$ For $t_c$, we list median, the 68\% and 95\% lower bounds and the
68\% and 95\% upper bounds.\\
$^f$ The number of degrees of freedom.}}

\begin{figure}[h]
\centerline{\leavevmode\epsfysize= 6 cm
\epsfbox{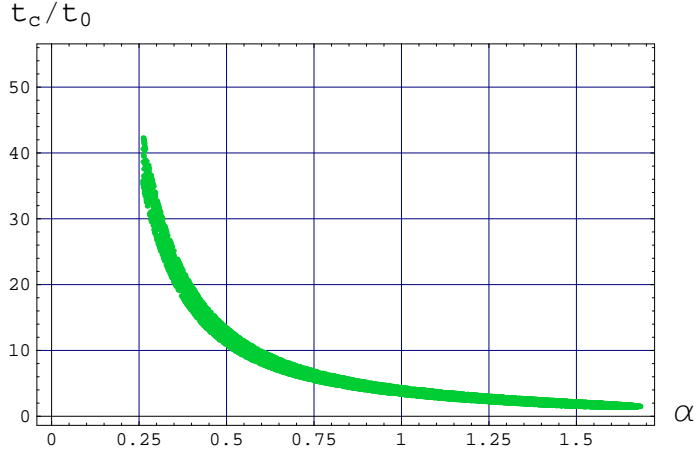}}
\caption[1]{\label{alphaT}\footnotesize%
The ratio of the collapse time from today, $t_c$, and the age of
the universe today, $t_0$, as a function of the linear potential
parameter $|\alpha |$. Fig.\ref{PdistrV2.eps}.}
\end{figure}

\newpage

Note that only the median, 68\% and 95\% confidence lower and upper limits of the
collapse time from today $t_c$ are given, since $\alpha=0$ is the
cosmological constant model with $t_c \rightarrow \infty$ (the
mean of $t_c$ is not well defined for this reason).
Computationally, we have to make a cutoff in $t_c$ for all models
that are longer lived than the computation limit. The dependence
of $t_c/t_0$ on $\alpha$ is shown in Fig.\ref{alphaT}, where $t_0$
is the age of the universe today.

\begin{figure}[h]
\vskip-0.1cm
\centerline{\epsffile{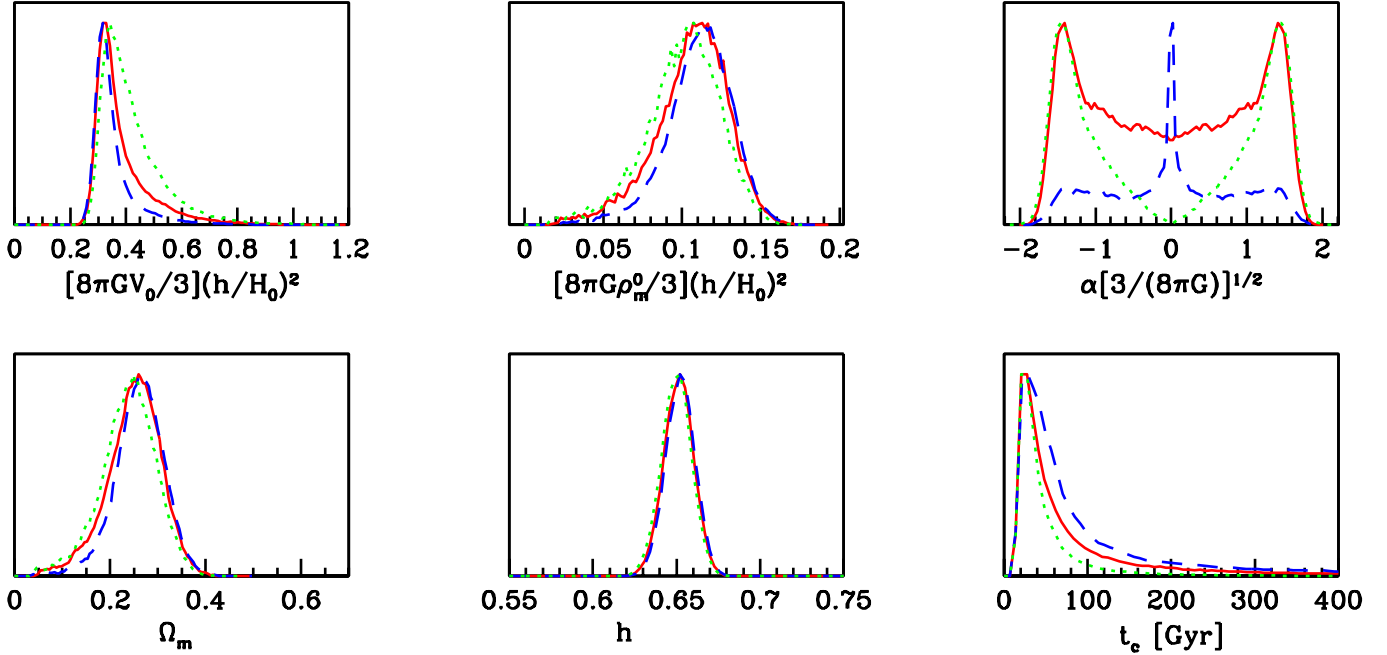}}
\vskip -10cm
\caption[1]{\label{pdfalprior}\footnotesize%
The effect of assuming different priors for $\alpha$, using only SN data (Riess
sample gold set, flux-averaged with $\Delta z=0.05$). The parameters are the
same as in Fig.\ref{pdf3set}. The solid, dashed, and dotted lines correspond to
priors of $p(\alpha) \propto 1$, $\alpha^{-0.5}$, and $\alpha$  respectively. }
\end{figure}
 Fig.\ref{pdfalprior} shows the effect of assuming different priors
for $\alpha$, using only SN data (Riess sample gold set,
flux-averaged with $\Delta z=0.05$). The parameters are the same
as in Fig.\ref{pdf3set}. The solid, dashed, and dotted lines
correspond to priors of $p(\alpha) \propto 1$, $\alpha^{-0.5}$,
and $\alpha$ respectively.
Table 2 shows how assuming different priors for $\alpha$ changes
the median, 68\% and 95\% lower and upper bounds on the collapse
time from today $t_c$.
\newpage
\begin{center}
{\footnotesize
{Table 2\\
Effect of assuming different priors for $\alpha$ on the collapse
time from today $t_c$ (in Gyrs)}\\
\begin{tabular}{|l|l|l|l|}
\hline
\hline
$P(\alpha) \propto $ & 1  & $\alpha^{-0.5}$ &  $\alpha$ \\
\hline
median & 56.01 & 96.92 & 35.04\\\hline
68\% lower bound & 35.75 & 38.77 &  27.41\\\hline
95\% lower bound & 19.18 & 19.38 &  17.91 \\\hline
68\% upper bound & 108.18 &  523.35 &  48.75\\\hline
95\% upper bound & 2704.01 & 714260.81 & 186.42\\
\hline
\hline
\end{tabular}
}
\end{center}

\section{MCMC versus Fisher Matrix}

 In this section we compare the joint confidence regions from our
MCMC calculations to that of the popular Fisher matrix estimates.
The Fisher matrix approach is an easy way to estimate the
confidence regions in an $n$-dimensional parameter space $p_i,$
($i=1,2, ..., n$), and it is especially powerful in the case
when the probability distributions are close to the Gaussian
distribution. The Fisher matrix is the inverse of the covariance
matrix in this $n$-dimensional parameter space, see \cite{Fisher}:
$$F \equiv C^{-1}. $$
 It is easy to calculate the covariance matrix $C$ using the MCMC data
 and marginalizing
 over additional parameters. 
 We will consider the distribution
 and confidence regions for $\alpha $ and $\Omega_m.$
\begin{figure}
\centerline{\leavevmode\epsfysize= 5.5cm
\epsfbox{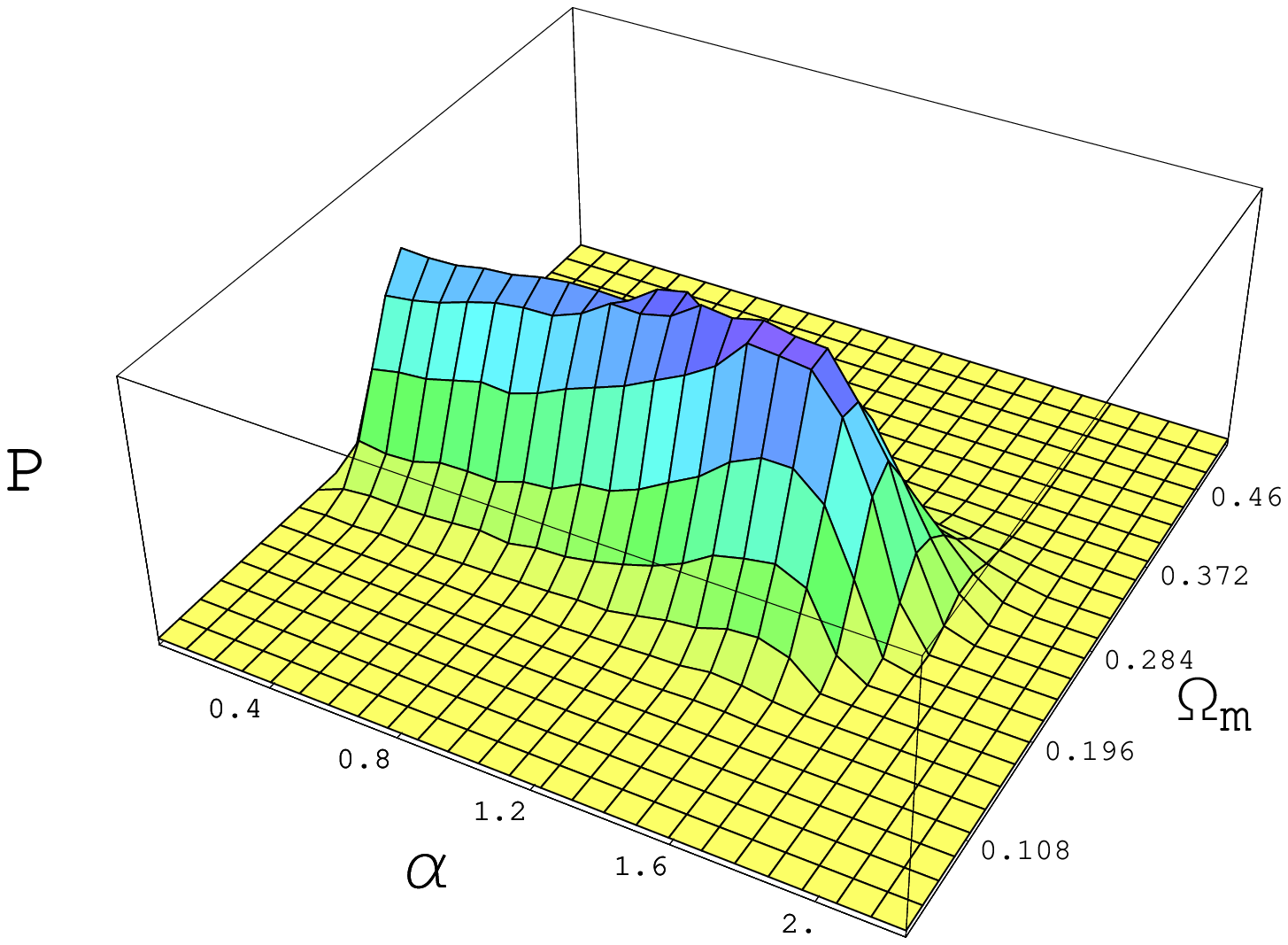} \leavevmode\epsfysize= 5.5cm
\epsfbox{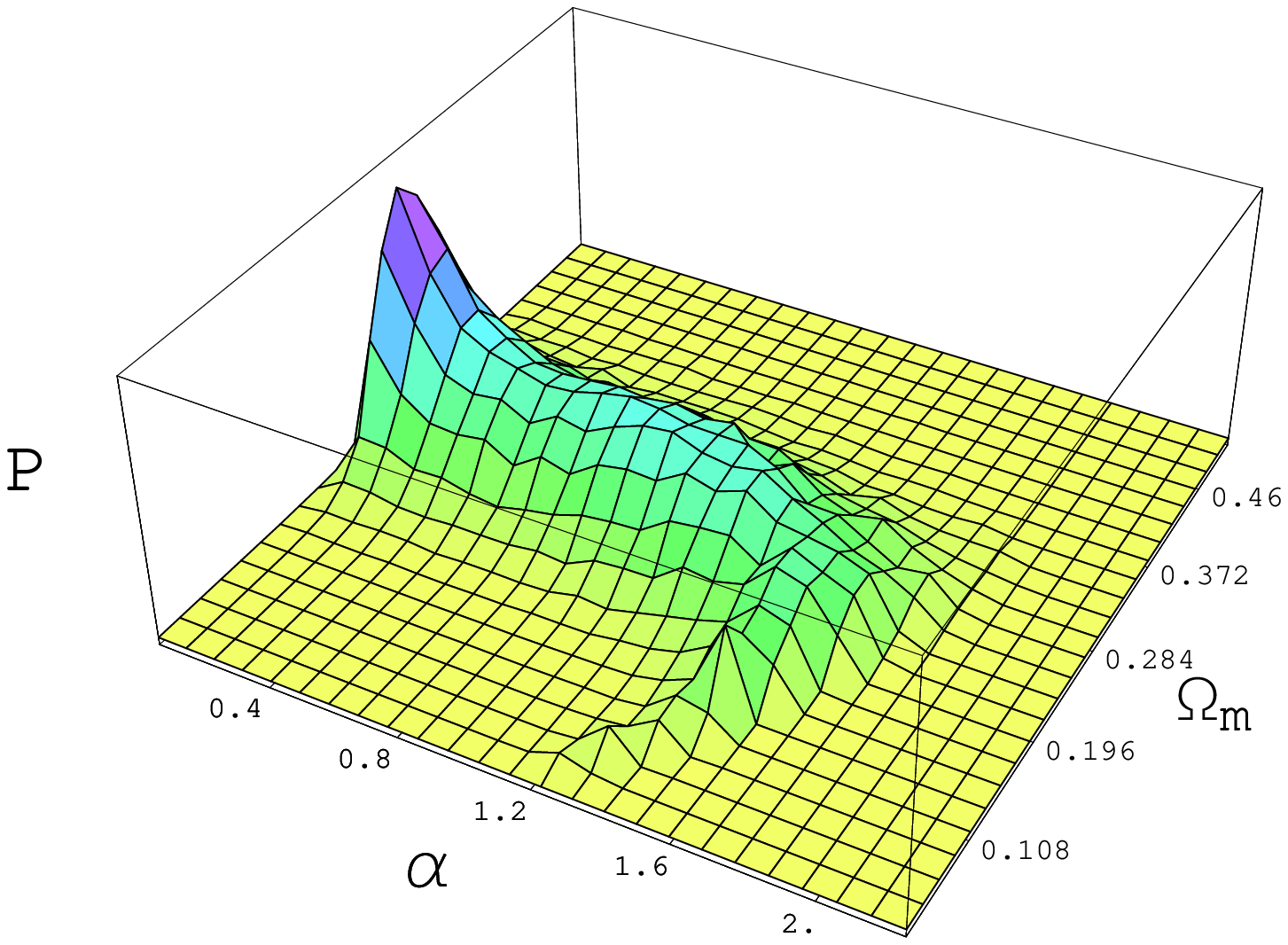}}
\caption[1]{\label{PdistrV2.eps}\footnotesize%
The probability distribution function for the set of parameters
$|\alpha|$ and $\Omega_m$ with uniform prior on $\alpha$ (left figure)
and with the prior proportional to $\alpha^{-0.5}$ (right figure).}
\end{figure}
In our case, the distribution of the parameter $\alpha $ (slope of
the potential) is significantly non-Gaussian (see
Fig.\ref{pdf3set}). There are a few facts that we have to take
into account. First, the symmetry of the picture is dictated by
the fact that the $-\alpha$ and $\alpha$ slopes in Eq.(1) are
indistinguishable. It also leads to a paradoxical statement that
$\Omega_m$ and $\alpha $ parameters become uncorrelated (the
corresponding off-diagonal elements of the Fisher matrix are 0).
Second, the actual parameter that we are trying to measure, and,
hopefully, distinguish from $ 0 $  (cosmological constant case) is
actually $|\alpha |.$  This means that in our estimations for
joint confidence regions we have to consider absolute value
$|\alpha |$ rather than $\alpha .$  In Fig.\ref{PdistrV2.eps} the
two-dimensional probability distribution functions for the set of
parameters $|\alpha|$ and $\Omega_m$  are presented for two
different cases - SN plus CMB and LSS data with uniform priors on
($\rho_m^0$, $V_0$, $\alpha$) (right plot) and SN data  with
$p(\alpha) \propto \alpha^{-0.5}$ (left plot). In both cases the
probability distributions are non-Gaussian.
\begin{figure}[b]
\hskip -3cm
\centerline{\leavevmode\epsfysize=9cm
\epsfbox{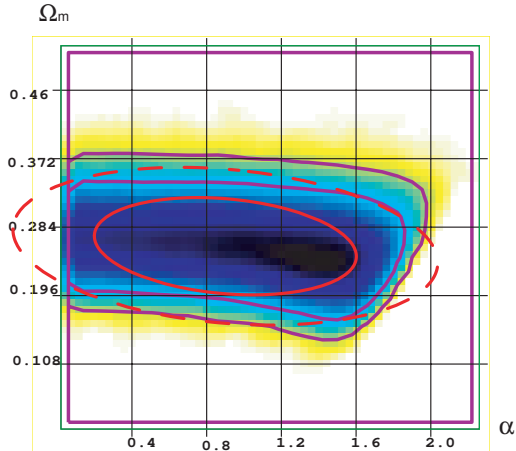}}
\vskip -3cm
\caption[1]{\label{MCfisher68Symmetric}\footnotesize%
The confidence region for the set of parameters $|\alpha|$ and
$\Omega_m$. The purple contours are MCMC $68 \%$ and $95 \%$
regions and red solid (dashed) ellipsoids correspond to $68 \% $ ($95 \%
$) Fisher matrix estimations. }
\end{figure}

The confidence regions form the Fisher matrix and MCMC method are
shown in  Fig.\ref{MCfisher68Symmetric}  for SN  with CMB and LSS
data and uniform priors on ($\rho_m^0$, $V_0$, $\alpha$) (left
plot in Fig.\ref{PdistrV2.eps}). The MCMC confidence regions are
presented by two uneven contours for 1-$\sigma$ (68\%) and
2-$\sigma$ (95\%) values. Two Fisher ellipses (red solid ellipse
corresponds to $68 \%$ confidence region and red dashed ellipse
corresponds to $95 \%$ region) are shown as well.
 It is clear that due to the non-Gaussian nature of the distribution the
 Fisher ellipses  could be used only for very rough estimations of the confidence
 regions. In fact, according to Fig.\ref{MCfisher68Symmetric},
$\alpha = 0$ (the cosmological constant case) would be excluded
at 1-$\sigma $ level for the Fisher matrix approach, that is
clearly not the case from the MCMC results. The  2-$\sigma$ regions
for both cases give similar results for $\alpha = 0$ and $10\% $
discrepancy for the upper bound on $ |\alpha |$.

\section{Predictions from Model-Independent Bounds on the Past Evolution}

We now derive complementary bounds on the cosmic doomsday by finding linear
potential models that fit completely under the 68\% and 95\% confidence
contours of the model-independent reconstruction of dark energy density
$\rho_X(z)$ from Ref.\cite{WangTegmark}.

The idea is to map out the boundaries of the past behavior of $\rho_X(z)$ in
a model-independent way based on SN Ia and other observations, in this
case by the means of splines. Splines have the advantage over polynomials, for
instance, that they do not have the tendency to wiggle in between measurement
points.
While the results will depend somewhat on the type of splines used, i.e.\ on
the number of parameters (the more parameters one uses, the more information is
extracted from the data, but the weaker the constraints will become), it is
important to realize how this approach differs from earlier, conventional,
model-dependent studies. The splines approach is able to model any arbitrary
function, not making any assumption about the functional form of $\rho_X(z)$ or
$w_X(z)$, contrary to the practice prevailing in present literature of using
two parameter fits.\footnote{Refs. \cite{WangTegmark} and \cite{Basset04} show
the perils of such practice.}

\begin{figure}[h]
\centerline{\leavevmode\epsfysize= 5 cm \epsfbox{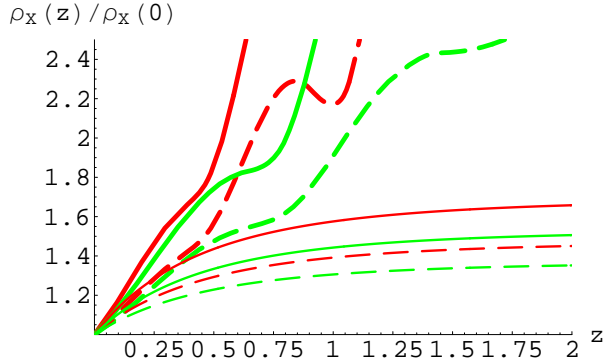}}
\caption[1]{\label{Model-independent Plot}\footnotesize%
The model-independent constraints on the past of $\rho_X(z)/\rho_X(0)$ and how
they constrain the linear model. The bounds are represented by thick lines,
while the corresponding linear model is shown by thin lines with same color and
same texture (same dashing). Green denotes the 3-parameter spline bound and its
linear model, while red shows the 4-parameter splines. The dashed lines depict
the 68\% confidence level, while the solid lines stand for 95\%. Since the
3-parameter constraint contours always lie inside the 4-parameter spline at a
given confidence level, one can argue that the 3-spline investigation is
sufficient for this analysis, yielding a minimal future lifetime of $20.6$ Gyr
at 95\% confidence level.}
\end{figure}

As a starting point for our study, we take the model-independent boundaries on
the past evolution of $\rho_X(z)/\rho_X(0)$ obtained in \cite{WangTegmark}.
Next, we solve the differential equations for the evolution history of the
universe with the linear potential for all possible initial conditions for the
scalar field consistent with the $95\%$ and $68\%$ confidence regions of
$\Omega_m$, and fit their corresponding past $\rho_X(z)/\rho_X(0)$ history into
the above envelope. The model parameters yielding the shortest lifetime, but
still not being in violation anywhere with this envelope, determine the minimal
lifetime obtainable by this method. These limiting cases, together with the
boundaries, are graphically represented in Fig.\ref{Model-independent Plot}.

Our approach is different from \cite{WangTegmark} in that we recognize that an
accurate prediction of the future can only be obtained by a model-specific
extrapolation. One can study the past model-independently, yet whatever the
conclusion from it is, the future will always depend on the specific underlying
physical model. Even in the absence of knowledge of the exact physics---as it
is the case presently with dark energy---one is forced to assume a specific
physical model to obtain a prediction on the future lifetime of the universe.
In our case, the model is the scalar field linear potential.

We use the Hubble time $H_0^{-1}$ as the time scale in our calculations. 
To translate the
lifetime into Gyr, we need the value of the Hubble constant $H_0$. The Hubble
constant, as obtained from the model-independent analysis in
\cite{WangTegmark},  has a mean value of $H_0=63.7 \rm{km/s/Mpc}$, corresponding
to a Hubble time $H_0^{-1} =15.36\rm{Gyr}$, for the 3-parameter splines, while
$H_0=63.5 \rm{km/s/Mpc}$ ($H_0^{-1} =15.41\rm{Gyr}$) for the 4-parameter spline.\footnote{These Hubble constant values are fully consistent with
other {\it direct} measurements. SN Ia data typically give a $H_0$ value
consistent with $H_0=60\pm 10\,$km$\,$s$^{-1}\,$Mpc$^{-1}$ \cite{Branch}.
The Hubble Key Project (based on Cepheid distances) 
gives $H_0=72\pm 8\,$km$\,$s$^{-1}\,$Mpc$^{-1}$ \cite{HST}.
Note that CMB data do {\it not} give a direct measurement of $H_0$.}

Even the $95\%$ error bars of the distribution for $h$ do not cause an uncertainty of more than $5.7\%$,
corresponding to $0.87\mathrm{Gyr}$, thus we will neglect this uncertainty in
our lifetime calculation in this section.

In this manner, we obtain for the future lifetime of the universe:

\begin{center}
{\footnotesize
{Table 3\\
The lower bound on future lifetime of the universe}\\
\ \\
\begin{tabular}{|l|c|c|} \hline
Spline Type & Confidence Level & Future Lifetime \\
\hline \hline
4-parameter & 95\% lower bound & 18.1 Gyr \\ 
 & $68\%$ lower bound & 22.2 Gyr \\ 
\hline
3-parameter & $95\%$ lower bound & 20.6 Gyr \\ 
 & $68\%$ lower bound & 25.3 Gyr \\ 
\hline
\end{tabular}
}
\end{center}

These results turn out to be a little weaker than our model-specific analysis
above. This is not unexpected; it will be very difficult, if not impossible, to
close the gap and raise the model-independent analysis to the same accuracy
level as the model-dependent one. After all, no assumption about the model in
the past is made.

The model-independent analysis also depends on the number of free parameters
used. The difference in results between the 3-parameter and the 4-parameter
spline is roughly $10\%$. As usual, stronger constraints are achieved by fewer
parameters at the cost of less flexibility. Since the 3-parameter spline
contour lies completely within the 4-parameter one, one can argue that for the
studied sample of supernovae using 3-parameter splines is sufficient.

We conclude that the results obtained by the model-independent study of the
past are somewhat less stringent than our results obtained in Sec.\ \ref{results}. Indeed, constraining the parameters of a given model directly is
the most precise way to proceed. However, it is quite convenient to have a
\emph{model-independent} analysis of the past, and then apply it to any
particular model for a quick estimate of the lifetime of the universe. This
method provides a useful shortcut, which lets one to obtain approximate results
without having to employ the whole machinery of MCMC that we have used to
obtain our results  in Sec.\ \ref{results}.

\section{Discussion and Summary}

We have studied a representative dark energy model with a linear potential,
$V(\phi)=V_0 (1 +\alpha \phi)$, assuming a flat universe.
This model has the interesting property of dooming the universe to a Big Crunch
for $\alpha  \neq 0$.
It is the simplest doomsday model, in which the universe
collapses rather quickly after it stops expanding.

Since this model has only one free parameter, $\alpha V_0$, it is well
constrained even by SN Ia data  alone (Riess sample), see Fig.\ref{pdf3set}.

We studied this model using current SN Ia (Riess sample), CMB (WMAP, CBI, and
ACBAR), and LSS (2dF) data. We have used flux-averaging to minimize the effect
of weak lensing \cite{Wang00b,WangMukherjee}, and included CMB and galaxy
survey data in a consistent manner \cite{WangTegmark}. We found that in the
context of the model with $V(\phi)=V_0 (1 +\alpha \phi)$, observations imply
that the cosmic doomsday is more than about 42 billion years from today at 68\%
confidence, and more than about 24 billion years from today at at the 95\%
confidence level.

These results represent a stronger bound on the doomsday time than the estimate
of $t_c \gtrsim 11$ Gyr based on the previously available
observational data \cite{Kalloshetal}. When the data from SNAP and Planck
satellites become available, it will be possible to further strengthen our bound
up to $t_c \gtrsim$ 40 Gyr at 95\% confidence \cite{Kalloshetal}.
Note that these timescales have been derived assuming a uniform prior on
$\alpha$ (the parameter of the linear model).

When one evaluates the cosmological consequences of a given
theory, one should specify not only its Lagrangian and the cosmological initial
conditions, but also the theoretical expectations (priors) for the values of the
parameters in the model. Since the change of the prior on the parameters is equivalent to changing the underlying physical model, there is nothing surprising in the dependence of the final results on the choice of the prior. For example, if one assumes, on the basis of some physical theory, that the cosmological constant is given by $|\Lambda| = e^{\gamma}$, and all values of $\gamma = \ln |\Lambda|$ are equally probable, one would come to a conclusion that the prior probability for  $\Lambda$ is strongly peaked at $\Lambda =0$. This would completely change not only the most probable value of dark energy density obtained from supernova experiments by using MCMC method, but also the famous anthropic bounds on $\Lambda$. However, in the absence of convincing theoretical arguments in favor of the uniform prior for $\ln \Lambda$, one usually assumes uniform or nearly uniform prior for $\Lambda$.  

Similarly, the main emphasis of our work was on the study of the linear model with the uniform prior for $\alpha$. To make our results more general and robust, we have also considered other physically motivated priors on $\alpha$ as
well, based on  \cite{Garriga:2003hj}. Changing the prior on $\alpha$
can change the bounds on the cosmic doomsday. We found that taking different
priors, $p(\alpha) \sim \alpha$ and $p(\alpha) \sim \alpha^{-0.5}$, altered the
95\% confidence lower bound on $t_c$ obtained for $p(\alpha) = const$ only by about 10\%. If one
assume the uniform prior for  $\ln |\alpha|$, one would conclude that the
universe will live for an exponentially long time \cite{GV,Garriga:2003hj}. However, if, just as in the case with the cosmological constant, one leaves theoretical speculations aside, one may want to concentrate on the results of our investigation with the uniform prior for $\alpha$.

Interestingly, we were able to obtain not only the lower bound on the doomsday
time, but the upper bound as well. For the uniform prior on $\alpha$  the full
bound on the doomsday time is $24\ {\rm Gyr} < t_c  < 3652$ Gyr at the 95\%
confidence level (SN Ia plus CMB and LSS data, see Table 1). 
In other words, if one believes that the dark energy is
described by the simplest linear model, and the parameter $\alpha$ can take any
value with equal probability, then our universe most probably will not collapse
earlier than in $24$ billion years, but it will most probably live no longer
than $3.65\times10^3$ billion years. This doomsday prediction is model-dependent: it is a consequence
of our assumption that dark energy is described by the linear model (1), and
that all values of the parameter $\alpha$ are equally probable. On the other
hand, as we already mentioned, this model is rather generic; it is the simplest
representative of a broad class of the dark energy models where the effective
cosmological constant can change continuously and the cosmological constant
problem can be solved using anthropic considerations
\cite{Linde,GV,Kallosh,Garriga:2003hj}. To improve our constraints on the
cosmic doomsday time in this class of models one would need further
cosmological observations and a deeper understanding of the theory
of dark energy.

\acknowledgements We thank Max Tegmark for helpful discussions. This work is
supported in part by NSF CAREER grant AST-0094335 (YW). The work by J.M.K.\ was
supported by the Stanford Graduate Fellowship  and the Sunburst Fund of the
Swiss Federal Institutes of Technology (ETH Zurich and EPF Lausanne). The work
by A.L.\ was supported by NSF grant PHY-0244728. The work by M.S.\ was
supported by DOE grant number DE-AC02-76SF00515.

\end{document}